# Probing QED and fundamental constants through laser spectroscopy of vibrational transitions in HD$^+$

J. Biesheuvel[1], J.-Ph. Karr[2,3], L. Hilico[2,3], K. S. E. Eikema[1], W. Ubachs[1] and J. C. J. Koelemeij[1]*

1. LaserLaB, Department of Physics and Astronomy, VU University, De Boelelaan 1081, 1081 HV Amsterdam, The Netherlands

2. Laboratoire Kastler Brossel, UPMC-Sorbonne Universités, CNRS, ENS-PSL Research University, Collège de France, 4 place Jussieu, 75005 Paris, France

3. Département de Physique, Université d'Evry Val d'Essonne, Boulevard François Mitterrand, 91025 Evry Cedex, France

Correspondence and requests for materials should be addressed to J.C.J.K. (email: j.c.j.koelemeij@vu.nl).

**The simplest molecules in nature, molecular hydrogen ions in the form of $H_2^+$ and $HD^+$, provide an important benchmark system for tests of quantum electrodynamics in complex forms of matter. Here, we report on such a test based on a frequency measurement of a vibrational overtone transition in $HD^+$ by laser spectroscopy. We find that the theoretical and experimental frequencies are equal to within 0.6(1.1) parts per billion, which represents the most stringent test of molecular theory so far. Our measurement not only confirms the validity of high-order quantum electrodynamics in molecules, but also enables the long predicted determination of the proton-to-electron mass ratio from a molecular system, as well as improved constraints on hypothetical fifth forces and compactified higher dimensions at the molecular scale. With the perspective of comparisons between theory and experiment at the 0.01 part-per-billion level, our work demonstrates the potential of molecular hydrogen ions as a probe of fundamental physical constants and laws.**

## Introduction

The possibility of accurate quantum electrodynamics (QED) calculations and the presence of narrow optical transitions between long-lived vibrational levels make laser spectroscopy of $H_2^+$ and $HD^+$ a sensitive tool to test fundamental physics at the molecular scale. Examples include tests of relativistic quantum mechanics and QED[1,2,3,4], and searches for physics beyond the Standard Model at the molecular scale[5,6] and beyond General Relativity[7,8]. Furthermore, spectroscopy of the molecular hydrogen ion has long been proposed as a means to determine the value of fundamental physical constants[1,2,4,9], and to realise precise molecular clocks[8,10].

Nearly all the above applications involve a comparison of state-of-the-art molecular theory with accurate results from (most often laser) spectroscopy. The most accurate results so far have been obtained using HD$^+$ ions stored in radiofrequency (r.f.) traps, sympathetically cooled by laser-cooled Be$^+$ ions[3,11], allowing tests at a relative uncertainty of a few parts per billion (p.p.b.) However, the most precise experimental result[3], a frequency measurement of the (*v*,*L*): (0,0) – (1,1) rovibrational line at 5.1 µm in HD$^+$ with a relative uncertainty of 1.1 p.p.b., was found to disagree by 2.7 p.p.b. with a more accurate theoretical value obtained from state-of-the-art *ab initio* molecular theory[1,2]. This disagreement has thus far been unresolved, and additional high-precision experimental data are needed to draw conclusions about the validity of the theoretical framework, and to open up the wide range of applications of molecular hydrogen ion spectroscopy mentioned above.

Here, we report on an optical frequency measurement of the (*v*,*L*): (0,2) – (8,3) vibrational overtone transition in HD$^+$ at 782 nm. Our experimental result is found to be in agreement with the theoretical prediction to within 0.6(1.1) p.p.b., thereby confirming the validity of relativistic quantum mechanics and QED in a vibrating molecular system at an unprecedented level. We subsequently exploit the agreement between theory and experiment for the first determination of the proton-electron mass ratio from a molecular system, and to put tighter constraints on the strength and range of 'fifth forces' at the molecular scale.

## Results

**Experimental procedure.** Our experimental apparatus and procedure for HD$^+$ spectroscopy are described in detail in the Methods section. In brief, trapped HD$^+$ molecular ions are cooled to ~10 mK by storing them together with laser-cooled Be$^+$ ions, and we excite and detect the ($v,L$): (0,2) – (8,3) transition by resonance-enhanced (1+1') multiphoton dissociation (REMPD) spectroscopy, see Fig. 1a. We acquire a spectrum by observing the loss of HD$^+$ due to REMPD, inferred from the change in the monitored Be$^+$ fluorescence induced by r.f. excitation of the HD$^+$ secular motion, for different values of the 782 nm spectroscopy laser frequency (Methods and Fig. 2a). The latter is stabilized and counted using an optical frequency comb laser (Methods).

**Line shape model for nonlinear least-squares fitting.** Hyperfine interactions lead to a manifold of 83 lines (Figs. 1b,c), the ~25 strongest of which are located within the range (-110 MHz, 140 MHz) around the hyperfine-less rovibrational frequency, $v_0$. Each hyperfine component is Doppler broadened to ~16 MHz, and the hyperfine structure is only partially resolved. Using a realistic spectral line shape function (Methods), we employ nonlinear least squares fitting to extract not only the transition frequency $v_{0,\text{fit}}$, but also the intensity of the 782 nm laser, $I_L$, the motional temperature of the HD$^+$ ions, $T_{\text{HD+}}$, and the temperature of the Be$^+$ ions during secular excitation, $T_i$, which are relevant parameters for the spectral analysis (see Methods). The 782 nm intensity we find agrees (within the fit error) with the intensity estimated from the laser beam waist and beam alignment uncertainty. Likewise, the fitted temperature agrees well with the results of molecular dynamics (MD) simulations.

The relevance of the additional fit parameters is illustrated by the correlation matrix, $\Omega$, of the estimated parameters, which reveals significant correlations between $v_{0,\text{fit}}$ and $T_{\text{HD+}}$, and $v_{0,\text{fit}}$ and $I_L$ (Fig. 2c). Adding fit parameters results in an increased error in the fitted value $v_{0,\text{fit}}$, which rises from 0.25 MHz (0.65 p.p.b. relative to $v_{0,\text{fit}}$) to 0.33 MHz (0.85 p.p.b.) when $T_i$, $T_{HD+}$, and $I_L$ are added as free fit parameters. Even so, the spectral fit result presented here marks the first time that a sub-p.p.b. resolution is achieved in molecular-ion spectroscopy.

**Systematic effects and frequency value.** The fit result $\nu_{0,fit}$ is corrected for various systematic frequency offsets due to electric and magnetic fields. We calculate the a.c. and d.c. Stark effect *ab initio* with high accuracy following the approach of Karr[10] to find the frequency shift due blackbody radiation (BBR), the r.f. trap field, and the electric fields of the lasers. The total Stark shift of -1.3(1) kHz is dominated by the shifts due to the 313 nm, 532 nm and 782 nm lasers (which are all on during excitation), with individual contributions of 0.008(1) kHz, -0.45(7) kHz, and -0.87(13) kHz, respectively. Here, the uncertainties of the laser beam intensities are the dominant source of frequency uncertainty. We also evaluate the Zeeman effect, which stems from the static quantization field of 0.19 mT directed along the z-axis of the trap (Methods). The polarization of the 782 nm laser is such that it induces $\sigma^-$ and $\sigma^+$ transitions at equal rates. The Zeeman effect leads to a shift to $\nu_{0,fit}$ by -16.9(3.2) kHz, the uncertainty of which is due to a possible 2% imbalance between the $\sigma^-$ and $\sigma^+$ rates caused by imperfect polarization optics and depolarization due to the windows of the vacuum chamber. Another uncertainty stems from the accuracy of the theoretical hyperfine structure[12], which enters through our spectral line shape function. We repeated the fit procedure with a spectral line shape function based on the hyperfine structure obtained with slightly altered spin coefficients (within their uncertainty range), for which we observe essentially no shift of $\nu_{0,fit}$. Compared to the 0.33 MHz statistical fit uncertainty of $\nu_{0,fit}$, the contribution of the above line shifts to the total frequency value and uncertainty is negligible, as are the frequency shifts due to the second-order Doppler effect[13] (<5 Hz) and the electric-quadrupole shift[14] (<0.1 kHz).

During REMPD, the motional dynamics in the laser-cooled Coulomb crystal are significantly influenced by several (laser-induced) chemical processes. For example, REMPD of an $HD^+$ ion leads to the formation of energetic fragments, and previously it was observed that the two reactions $HD^+ + h\nu_1 + h\nu_2 \rightarrow H(1s) + D^+$ + k.e., and $HD^+ + h\nu_1 + h\nu_2 \rightarrow D(1s) + H^+$ + k.e. occur with about equal probability (here k.e. stands for kinetic energy, and $\nu_1$ and $\nu_2$ for the frequencies of the 782 nm and 532 nm lasers, respectively)[15]. Due to their relatively high charge-to-mass ratio, the produced protons are not stably trapped in our apparatus. Indeed, we observe only cold, trapped deuterons after REMPD, indicating that each deuteron transferred most of its 0.41 eV kinetic energy to the ions in the Coulomb crystal (which itself contains only 2 meV of thermal energy at 10 mK), causing (at least transiently) significant heating of the crystal. Other reactions with energetic ionic products are the result of collisions with $H_2$ molecules in the background vapour, and we also take the reactions $Be^+(^2P_{3/2}) + H_2 \rightarrow BeH^+ + H(1s)$ + k.e., $HD^+ + H_2 \rightarrow H_2D^+ + H(1s)$ + k.e., and $HD^+ + H_2 \rightarrow H_3^+ + D(1s)$ + k.e. (with ionic product kinetic energies of 0.25, 0.36, and 0.66 eV, respectively) into account here. We determine reaction rates from measured loss rates of $Be^+$ and $HD^+$, which are in agreement with expected Langevin reaction rates given the pressure of $1\times10^{-8}$ Pa in our vacuum apparatus[16]. We include these processes in realistic MD simulations (Methods) employing leapfrog integration with an adaptive time step, to ensure that collisions between energetic ions are correctly handled. Under these conditions, our MD simulations reveal average $HD^+$ z-velocity distributions which deviate significantly from thermal (Gaussian) distributions, an effect which has hitherto been neglected by the widespread assumption that laser-cooled ion ensembles exhibit thermal velocity distributions. As shown in Fig. 3, we empirically find that the velocity distributions are better represented by *q*-Gaussians[17], which have the additional advantage that the effect of a given chemical reaction can be parameterized by a corresponding *q*-value (with *q*=1 corresponding to a Gaussian distribution). Taking all reactions and reaction rates into account, we find that the velocity distribution is best characterized by a *q*-Gaussian with *q* ranging between 1.00 and 1.07, depending on the assumed thermalisation rate and the REMPD rate. We therefore include *q*-Gaussians in our spectral line shape function (Methods). Comparing scenarios with minimum and maximum expected thermalisation rates (corresponding to *q*=1.07 and *q*=1.00, respectively) we find a mean shift of -0.25(25) MHz with respect to the case of *q*=1.00. The origin of this shift lies in the overlap and saturation of Doppler-broadened hyperfine components in the spectrum (Fig. 2d). A similar frequency shift may occur when micromotion sidebands are present. Here we make a distinction between excess radial micromotion caused by radial static offset fields, and axial micromotion which could arise from an axial r.f. field due to geometrical imperfections of our ion trap. A finite-element analysis of the trap's electric field reveals that such an axial field will be approximately constant over the extent of the Coulomb crystal, which implies that the corresponding micromotion contribution to the line shape will be homogeneous. We use the fluorescence correlation method of Berkeland *et al.*[13] to find a small micromotion amplitude of 11(4) nm amplitude along the 782 nm laser beam as follows. Measurements of the r.f. field amplitude are performed on the $Be^+$ ions using the 313 nm laser beam (which co-propagates with the 782 nm laser beam). Repeating such measurements for various values of the trap r.f. voltage allows separating the axial and the radial component (the latter being due to the residual projection of the laser beam onto the radial direction). We find that the axial micromotion component is dominant, which is explained as follows. Firstly, the 782 nm laser is aligned parallel to the trap axis, so that the radial micromotion amplitude is suppressed by the near-zero angle between the 782 nm laser and the radial directions. Secondly, since the surrounding Be$^+$ ions shield the HD$^+$ from static radial offset fields, the HD$^+$ ions are trapped symmetrically about (and close to) the nodal line of the r.f. field. These two conditions limit the contribution of the radial micromotion to well below the measured axial amplitude of 11(4) nm. The resulting sidebands are included in the spectral line shape function, leading to an additional frequency shift of -55(20) kHz with respect to the case of zero micromotion. After correcting for all systematic frequency shifts (Table 1), we finally obtain $\nu_0$= 383,407,177.38(41) MHz.

## Discussion

Our frequency value $\nu_0$ of the (*v*,*L*): (0,2) – (8,3) transition is in good agreement with the more accurate theoretical value[2] $\nu_{0,theo}$= 383,407,177.150(15) MHz. The contribution of the QED terms[1,2,18] to this frequency amounts to -1547.225(15) MHz (-4035.46(4) p.p.b.), and our measurement therefore confirms the validity of QED in a molecular system at an unprecedented level of 2.7×10$^{-4}$. We furthermore note that our measurement is sensitive to (and in agreement with) QED terms up to order $m_e \alpha^7$, given that the $m_e \alpha^7$ term contributes 780(15) kHz (*i.e.* 2.03(4) p.p.b.) to the transition frequency.

Salumbides *et al.*[5] showed that spectroscopic tests of molecular QED can be used to set upper bounds on a hypothetical 'fifth force', acting between hadrons at separations of the order of 1 Å, and arising from the exchange of unknown, massive virtual particles. Modelling such a hadron-hadron interaction with a Yukawa-type potential of the form $\hbar c\, \alpha_5\, e^{-r/\lambda}/r$ and computing the resulting frequency shift to $v_0$, we can exploit the 1 p.p.b. agreement between theory and experiment to rule out (at the 90% confidence level) interactions at a range $\lambda$ = 1-2 Å (corresponding to force-carrying particles with mass $m_Y$ = 1-2 keV$c^{-2}$) with an interaction strength relative to the fine-structure constant, $|\alpha_5/\alpha|$, larger than 5-8×10$^{-10}$ (Fig. 4). In a similar way, applying the Arkani-Hamed-Dimopoulos-Dvali formulation to probe the effect of compactified higher dimensions on energy levels in molecules[6], we place an improved upper bound of 0.5 µm on the compactification radius of higher dimensions in eleven-dimensional M-theory.

Four decades ago, Wing *et al.*[4] suggested that molecular theory could one day be used to translate measured vibrational frequencies of HD$^+$ to a new value of the proton-electron mass ratio, $\mu$. The high accuracy of our result and the good agreement with theory, which assumes the 2010 Committee on Data for Science and Technology (CODATA) recommended value $\mu_{CODATA10}$, now allow the first determination of $\mu$ from molecular vibrations. Previously[19] we had derived the sensitivity relation $\delta\mu/\mu$ = -2.66 $\delta v_0/v_0$, which we employ to adjust $\mu_{CODATA10}$ to a new value, $\mu_{HD+}$, such that the theoretical frequency matches our experimental value. We thus find $\mu_{HD+}$=1,836.1526695(53), with a relative uncertainty of 2.9 p.p.b. which approaches that of the values taken into account in the 2010 CODATA adjustment[20]. For example, the value reported by Farnham *et al.* is only a factor of 1.3 more precise than our result[21]. While the precision of a very recent determination by Sturm *et al.* is still 31 times higher[22], we point out that our method yields $\mu$ as a single parameter from molecular vibrations, whereas most other $\mu$ values are ratios of individual determinations of the electron and proton relative atomic masses (exceptions are $\mu$ determinations from atomic laser spectroscopy[23,24]). Therefore, the agreement of $\mu_{HD+}$ with all other values of $\mu$ forms an additional consistency check of the various (and vastly different) methods used (Fig. 5b). In particular, it implies that relativistic quantum mechanics and QED consistently describe at the few-p.p.b. level such diverse systems as the bound electron[22,25,26], antiprotonic helium[23], and the molecular hydrogen ion. Furthermore, of all the methods which produce $\mu$ as a single parameter, our approach is surpassed only by spectroscopy of antiprotonic helium, which is 2.3 times more precise but additionally requires charge, parity and time reversal invariance[23] (Fig. 5a).

In principle, the transition frequency $\nu_0$ depends on the value of other fundamental constants as well, such as the deuteron-proton mass ratio[7], $m_d/m_p$, the fine structure constant, and the proton charge radius[1,2]. However, the sensitivity of $\nu_0$ to changes in $\mu$ is known to be three times larger than the second largest one, the $m_d/m_p$ sensitivity[7]. Moreover, if we propagate the uncertainties of the 2010 CODATA values of the fundamental constants through the sensitivity relations, we find that the relative contributions by $\mu$, $m_d/m_p$, $\alpha$, and the proton radius to the theoretical uncertainty of $\nu_0$ are 154, 11.6, 0.004, and 5.13 parts per trillion, respectively. We therefore conclude that $\mu$ is the correct parameter to constrain.

The error budget in Table 1 reveals that the experimental uncertainty is limited by Doppler broadening. To overcome this, more involved Doppler-free two-photon spectroscopy of HD$^+$ has been proposed[9], which should reduce the uncertainty to below $1\times10^{-12}$. This should enable a comparison between theory and experiment which would not only test the QED description of chemically bonded particles at an unprecedented level, but also produce a new value of $\mu$ with an uncertainty below 0.1 p.p.b., surpassing the most precise determination of $\mu$ obtained from independent electron and proton relative atomic mass measurements[22], which represent a completely different method (Fig. 5c). Our work demonstrates the potential of molecular hydrogen ions for the determination of mass ratios of fundamental particles, as well as stringent tests of QED, and searches for new physics.

**Methods**

**Experimental procedure.** We typically store 40 to 85 HD$^+$ ions in a linear rf trap, together with about 750 Be$^+$ ions, which are laser cooled to a temperature of 5-10 mK using a continuous-wave (CW) 313 nm laser beam propagating along the symmetry (*z*) axis of the trap. As a consequence, only the axial motion of the Be$^+$ ions is laser-cooled directly. However, the three-dimensional extent of the Coulomb crystal ensures good coupling of the axial motion to the radial motion of the Be$^+$ and HD$^+$ ions, so that both Be$^+$ and HD$^+$ are efficiently cooled in all three dimensions. Although larger numbers of ions may be trapped, smaller ion numbers ensure better reproducibility of the experimental conditions. The stronger confinement of HD$^+$ by the pseudopotential leads to the formation of a string or zig-zag structure of HD$^+$ ions along the nodal line of the r.f. field, which coincides with the *z*-axis. Be$^+$ ions arrange themselves in a three-dimensional ellipsoidal structure surrounding the HD$^+$. At temperatures below 0.1 K, the two-species ion ensemble solidifies into a Coulomb crystal. 313 nm fluorescence photons emitted by Be$^+$ are imaged onto an electron multiplied charge-coupled device (EMCCD) camera and a photomultiplier tube. We obtain a measure of the number of trapped HD$^+$ ions by resonantly driving their radial secular motion (~800 kHz) using an a.c. electric field[19]. MD simulations indicate that this 'secular excitation' heats up and melts the ion ensemble, heating the Be$^+$ ions to an average temperature (in the *z*-direction) $T_{Be+,av}$ proportional to the number of HD$^+$, $N_{HD+}$. Typically, $T_{Be+,av}$ = 2-4 K. For a fixed cooling-laser detuning $\Delta$=-18 $\Gamma$ (with $\Gamma$=2$\pi$×19.4 MHz the natural linewidth of the 313 nm cooling transition), this temperature rise leads to Doppler broadening and, thus, to a significant rise in the 313 nm fluorescence rate with average value *F*. Whereas previous work approximated the fluorescence rise versus $N_{HD+}$ by a linear dependence[27], we realistically model the (nonlinear) dependence of $F$ on $N_{HD+}$, and take this into account in our analysis (see Spectral line shape function for fitting).

To excite the ($v,L$): (0,2) – (8,3) rovibrational transition at 782 nm, we send a CW 782 nm laser beam along the $z$-axis, counter-propagating the 313 nm laser beam. The 782 nm radiation is obtained from a titanium:sapphire laser with a linewidth of 0.5 MHz, which is frequency locked to an optical frequency comb laser as follows. The frequency of the optical beat note of the 782 nm laser with a nearby mode of the comb is measured by a counter every 30 ms. From the measured beat-note frequency, the comb repetition rate, comb carrier-envelope offset frequency, and the comb mode number we determine the laser's optical frequency. The counter output is fed into a digital feedback loop, which controls the 782 nm laser so as to stabilise the beat-note frequency to a pre-set value. The comb is fully referenced to a GPS-disciplined rubidium atomic clock (providing long-term relative accuracy at the level of $2\times10^{-12}$), and the frequency uncertainty of our optical frequency measurement system (10 kHz) is limited solely by the frequency instability of the locked 782 nm laser averaged over the 10 s of REMPD. To detect excitation to the $v$=8 state by the 782 nm laser, we overlap this beam with a CW 532 nm laser beam, leading to REMPD of HD$^+$ (Fig. 1a). An experimental cycle looks as follows. We first load a fresh sample of HD$^+$ and find a measure, $F_i$, of the initial ion number, $N_i$, by secular excitation. We subsequently lower the 313 nm cooling laser intensity and detuning to reduce the ion temperature to ~10 mK, and expose the ions to the 313 nm, 782 nm and 532 nm lasers for 10 s. Afterwards, we apply secular excitation to determine the fluorescence level $F_f$ corresponding to the remaining ion number, $N_f$, and we define a signal $S = (F_i - F_f)/F_i$. Repeating this cycle for different values of the 782 nm laser frequency, $v$, we obtain a spectrum consisting of 1772 points (Fig. 2a).

Noise in the spectrum is primarily due to the stochastic nature of REMPD of small $HD^+$ ensembles with a mean rotational-hyperfine occupancy of the order of one ion per state.

**Spectral line shape function for fitting.** Hyperfine rate equations allow computing the evolution of a sample of $HD^+$ ions, with an initial thermal rotational distribution[28] corresponding to $T$ =300(5) K, under the influence of REMPD lasers, BBR and losses due to chemical reactions with background-gas molecules (Figs. 1b,d). Population in rotational states with $L$=0-5 is included, thus ignoring the 2.4% population in $L$=6 and higher. Accurate hyperfine line strengths at the magnetic sub-state level are obtained by extending the approach of Koelemeij[29] so as to include hyperfine structure and the Zeeman effect. The response to the 782 nm laser of each hyperfine level is modelled using a Doppler-broadened profile based on $q$-Gaussians. For the assessment of the Zeeman effect, the hyperfine rate equations are adapted to include Zeeman line shifts and broadening due to hyperfine line splitting (which remains much smaller than the 16 MHz Doppler width). The $q$ parameter is determined from realistic MD simulations (which take into account the time-dependent trap potential and momentum changes by scattering of photons from the cooling laser) with an uncertainty limited by the minimum and maximum expected rates of heating events. Solving the rate equations allows predicting the fractional loss of $HD^+$, $\gamma$, as a function of the 782 nm laser frequency relative to the hyperfine-less frequency, $\nu-\nu_0$, the intensity, $I_L$, of the 782 nm laser, and the $HD^+$ motional temperature, $T_{HD+}$. We obtain a smooth, continuous spectral line shape function for fitting as follows. First, we compute $\gamma$ on a dense three-dimensional grid of values ($\nu$, $I_L$, $T_{HD+}$), which we subsequently interpolate to obtain a four-dimensional function $\gamma(\nu-\nu_{0,fit}, I_L, T_{HD+})$. Using the shorthand notation $\gamma(\nu-\nu_{0,fit}, I_L, T_{HD+})=\gamma$, and making use of the linear dependence of $T_{Be+,av}$ on $N_{HD+}$ found from MD simulations, the average Be$^+$ temperatures during secular excitation of HD$^+$ before and after REMPD are given by $T_i \equiv T_{Be+,av}(N_i)$ and $T_f \equiv T_{Be+,av}((1-\gamma)N_i)$, respectively. We consecutively insert these temperature values in the scattering rate function $g$, defined as

$$g(T_{Be^+}) = \frac{\Gamma}{2}\sqrt{\frac{m_{Be^+}}{2\pi k_B T_{Be^+}}} \int_{-\infty}^{\infty} \frac{s}{1+s+4(\Delta - kv_k)^2/\Gamma^2} \exp\left(-\frac{mv_k^2}{2k_B T_{Be^+}}\right) dv_k, \quad (1)$$

which includes the beryllium mass, $m_{Be+}$, and the fixed 313 nm laser detuning, wave vector, **k**, and saturation parameter, $s$. The integration is performed over the distribution of Be$^+$ velocities **v** along **k**, i.e. $v_k \equiv \mathbf{k}\cdot\mathbf{v}/|\mathbf{k}|$. This function is used to model the fluorescence rise $F$ observed in the experiment, and allows us to construct a five-dimensional fit function $G(\nu-\nu_{0,fit}, I_L, T_{HD+}, T_i) \equiv [g(T_i) - g(T_f)]/[g(T_i) - g_0]$, with $g_0$ the steady-state scattering rate before secular excitation. The function $G$ realistically models the signal $S$ (Fig. 2c), and is used for fitting.

We employ the above model also to estimate the uncertainties due to several systematic effects. For example, MD simulations predict a slight increase of $T_{HD+}$ and $q$ with increased REMPD rates in the scenario of inefficient thermalisation of fast D$^+$. We can estimate the frequency shift due to this effect by making both $T_{HD+}$ and $q$ REMPD-rate dependent. This leads to an additional frequency uncertainty of 61 kHz, which is included in the error associated with Doppler effects due to chemistry (Table 1). From our model we furthermore deduce that ignoring states with $L \geq 6$ introduces an uncertainty of 32 kHz, while the possible 5 K error in the BBR temperature estimate (which takes into account day-to-day temperature variations and the possibility that the trap electrodes may be at a slightly higher temperature due to r.f. dissipation) translates to a 5 kHz effect.

**References**

1. Korobov, V. I., Hilico, L. & Karr, J.-Ph. $m\alpha^7$-Order corrections in the hydrogen molecular ions and antiprotonic helium. *Phys. Rev. Lett.* **112,** 103003 (2014).

2. Korobov, V. I., Hilico, L. & Karr, J.-Ph. Theoretical transition frequencies beyond 0.1 p.p.b. accuracy in $H_2^+$, $HD^+$, and antiprotonic helium. *Phys. Rev. A* **89,** 032511 (2014).

3. Bressel, U. *et al.* Manipulation of individual hyperfine states in cold trapped molecular ions and application to $HD^+$ frequency metrology. *Phys. Rev. Lett.* **108,** 183003 (2012).

4. Wing, W. H., Ruff, G. A., Lamb, Jr., W. E. & Spezeski, J. J. Observation of the infrared spectrum of the hydrogen molecular ion $HD^+$. *Phys. Rev. Lett.* **36,** 1488-1491 (1976).

5. Salumbides, E. J. *et al.* Bounds on fifth forces from precision measurements on molecules. *Phys. Rev. D* **87,** 112008 (2013).

6. Salumbides, E. J., Schellekens, A. N., Gato-Rivera, B. & Ubachs, W. Constraints on extra dimensions from precision molecular spectroscopy. *New. J. Phys.* **17,** 033015 (2015).

7. Schiller, S. & Korobov, V. Tests of time independence of the electron and nuclear masses with ultracold molecules. *Phys. Rev. A* **71,** 032505, (2005).

8. Schiller, S., Bakalov, D. & Korobov, V. I. Simplest molecules as candidates for precise optical clocks. *Phys. Rev. Lett.* **113,** 023004 (2014).

9. Tran, V. Q., Karr, J.-Ph., Douillet, A., Koelemeij, J. C. J. & Hilico, L. Two-photon spectroscopy of trapped $HD^+$ ions in the Lamb-Dicke regime. *Phys. Rev. A* **88,** 033421 (2013).

10. Karr, J.-Ph. $H_2^+$ and $HD^+$: candidates for a molecular clock, *J. Mol. Spectrosc.* **300,** 37-43 (2014).

11. Koelemeij, J. C. J., Roth, B., Wicht, A., Ernsting, I. & Schiller, S. Vibrational spectroscopy of $HD^+$ with 2-p.p.b. accuracy. *Phys. Rev. Lett.* **98,** 173002 (2007).

12. Bakalov, D., Korobov, V. I. & Schiller, S. High-precision calculation of the hyperfine structure of the $HD^+$ ion. *Phys. Rev. Lett.* **97,** 243001 (2006).

13. Berkeland, D. J., Miller, J. D., Bergquist, J. C., Itano, W. M. & Wineland, D. J. Minimization of ion micromotion in a Paul trap. *J. Appl. Phys.* **83**, 5025-5033 (1998).

14. Bakalov, D. & Schiller, S. The electric quadrupole moment of molecular hydrogen ions and their potential for a molecular ion clock. *Appl. Phys. B* **114,** 213-230 (2014).

15. Roth, B., Koelemeij, J. C. J., Daerr, H. & Schiller, S. Rovibrational spectroscopy of trapped molecular hydrogen ions at millikelvin temperatures. *Phys. Rev. A* **74,** 040501(R) (2006).

16. Wineland, D. J. *et al.* Experimental Issues in Coherent Quantum-State Manipulation of Trapped Atomic Ions. *J. Res. Natl. Inst. Stand. Technol.* **103,** 259 (1998).

17. Umarov, S., Tsallis, C. & Steinberg, S. On a *q*-central limit theorem consistent with nonextensive statistical mechanics. *Milan J. Math.* **76,** 307–328 (2008).

18. Korobov, V. I. Relativistic corrections of $m\alpha^6$ order to the rovibrational spectrum of $H_2^+$ and $HD^+$ molecular ions. *Phys. Rev. A* **77,** 022509 (2008).

19. Koelemeij, J. C. J., Noom, D. W. E., De Jong, D., Haddad, M. A. & Ubachs, W. Observation of the *v'*=8 - *v*=0 vibrational overtone in cold trapped $HD^+$. *Appl. Phys. B.* **107,** 1075-1085 (2012).

20. Mohr, P. J., Taylor, B. N. & Newell, D. B. CODATA recommended values of the fundamental physical constants: 2010. *Rev. Mod. Phys.* **84,** 1527-1605 (2012).

21. Farnham, D. L., Van Dyck, Jr., R. S. & Schwinberg, P. B. Determination of the electron's atomic mass and the proton/electron mass ratio via Penning trap mass spectroscopy. *Phys. Rev. Lett.* **75,** 3598-3601 (1995).

22. Sturm, S. *et al.* High-precision measurement of the atomic mass of the electron. *Nature* **506,** 467-470 (2014).

23. Hori, M. *et al.* Two-photon laser spectroscopy of antiprotonic helium and the antiproton-to-electron mass ratio. *Nature* **475,** 484-488 (2011).

24. De Beauvoir, B. *et al.* Absolute frequency measurement of the 2S-8S/D transitions in hydrogen and deuterium: new determination of the Rydberg constant. *Phys. Rev. Lett.* **78,** 440-443 (1997).

25. Beier, T. *et al.* New determination of the electron's mass. *Phys. Rev. Lett.* **88,** 011603 (2002).

26. Verdu., J. *et al.* Electronic g factor of hydrogenlike oxygen $^{16}O^{7+}$. *Phys. Rev. Lett.* **92,** 093002 (2004).

27. Schneider, T., Roth, B., Duncker, H., Ernsting, I. & Schiller, S. All-optical preparation of molecular ions in the rovibrational ground state. *Nature Phys.* **6,** 275-278 (2010).

28. Koelemeij, J. C. J., Roth, B. & Schiller, S. Blackbody thermometry with cold molecular ions and application to ion-based frequency standards. *Phys. Rev. A* **76,** 023413 (2007).

29. Koelemeij, J. C. J. Infrared dynamic polarizability of HD$^+$ rovibrational states. *Phys. Chem. Chem. Phys.* **13,** 18844 (2011).

30. Mohr, P. J., Taylor, B. N. & Newell, D. B. CODATA recommended values of the fundamental physical constants: 2006. *Rev. Mod. Phys.* **80,** 633-730 (2008).

**Acknowledgements** We are indebted to J. Bouma, T. Pinkert and R. Kortekaas for technical assistance, and to V. Korobov, E. Hudson and R. Gerritsma for fruitful discussions. We thank E. Salumbides for providing the curve for $D_2$ in Fig. 4. This research was funded through the Netherlands Foundation for Fundamental Research on Matter (FOM), the COST action MP1001 IOTA, and the Dutch-French bilateral Van Gogh Programme. SURFsara ([www.surfsara.nl](www.surfsara.nl)) is acknowledged for the support in using the Lisa Compute Cluster for MD simulations. J.C.J.K. thanks the Netherlands Organisation for Scientific Research (NWO) and the Netherlands Technology Foundation (STW) for support.

**Author contributions** J.B. and J.C.J.K built the apparatus, and acquired and analysed data. K.S.E.E. provided the optical frequency measurement system. J.-Ph.K., L.H., J.B. and J.C.J.K. evaluated systematic frequency shifts and performed simulations. W.U. and J.C.J.K. conceived and supervised the experiment. J.B. and J.C.J.K. wrote the manuscript, which was edited and approved by all co-authors.

**Competing financial interests** The authors declare no competing financial interests.

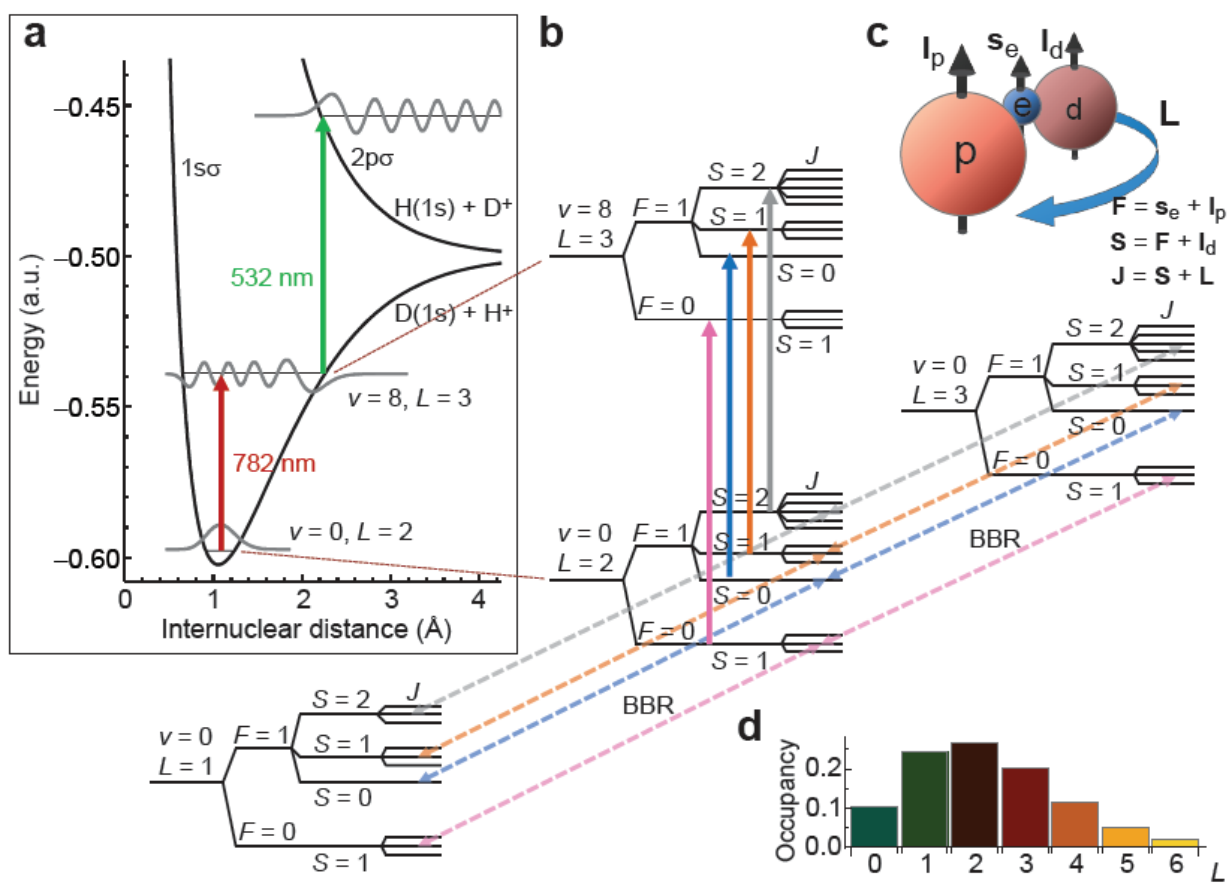

**Figure 1| Partial level diagram, excitation scheme and interaction with blackbody radiation.** The ($v,L$):(0,2)-(8,3) rovibrational transition is excited and detected through REMPD by 782 nm and 532 nm laser radiation (**a**, solid arrows), which leads to the loss of HD$^+$ by dissociation into either the H(1$s$) + D$^+$ or the D(1$s$) + H$^+$ channel[15]. Magnetic interactions between the proton spin $\mathbf{I}_p$, deuteron spin, $\mathbf{I}_d$, electron spin, $\mathbf{s}_e$, and molecular rotation, $\mathbf{L}$, lead to hyperfine structure in the excitation spectrum (**b, c**). Blackbody radiation induces rotational transitions (**b,** dashed arrows) between $v$=0, $L$=2 and states with $L$=1,3 which are in turn coupled to $L$=0,4 (not shown), and the population in each hyperfine state depends on the balance between the REMPD rate and the rate at which BBR replenishes population. These effects are included in a hyperfine rate equation model which is used to obtain a realistic spectral line shape function. Treating this at the level of individual

hyperfine states naturally subdivides the HD$^+$ population into different spin configuration classes, which are only weakly coupled by the electric dipole transitions induced by the lasers and BBR, and therefore evolve independently during REMPD. **d,** Rotational distribution of HD$^+$ ($v$=0) at an ambient BBR temperature of 300 K. The population of the $v$=0, $L$=2 initial state amounts to 27%.

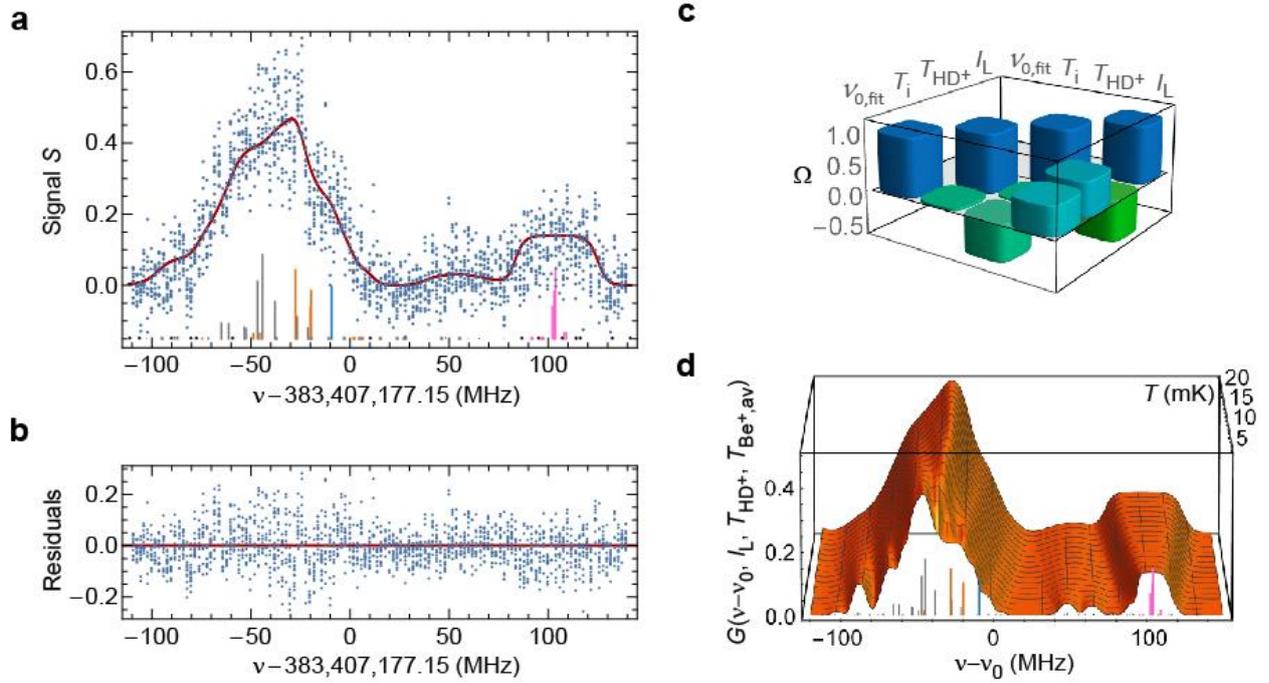

**Figure 2| Spectral line shape function and fit to REMPD spectrum. a**, Measured dimensionless REMPD signal $S$ versus 782 nm laser frequency (blue dots) and fitted line shape function (red curve). The underlying individual hyperfine components are also shown, following the same colour coding as used in Fig. 1b. The hyperfine components are offset vertically for clarity. **b,** Fit residuals versus 782 nm laser frequency (blue dots). **c**, The correlation matrix, $\Omega$, reveals correlations between $\nu_{0,\text{fit}}$ and the other fitted parameters, and the uncertainty in the fitted value $\nu_{0,\text{fit}}$ rises from 0.25 MHz (0.65 p.p.b. relative to $\nu_{0,\text{fit}}$) to 0.33 MHz (0.85 p.p.b.) when $T_i$, $T_{\text{HD}^+}$, and $I_L$ are added as free fit parameters. **d**, Dimensionless spectral line shape function $G$ obtained with the rate equation model (Methods) assuming $I_L = 1\times10^7$ W/m$^2$ and $T_{\text{Be}^+,\text{av}}$ = 4 K, showing the non-trivial combined effect of Doppler broadening, line overlap and saturation versus HD$^+$ temperature (axis shown in perspective).

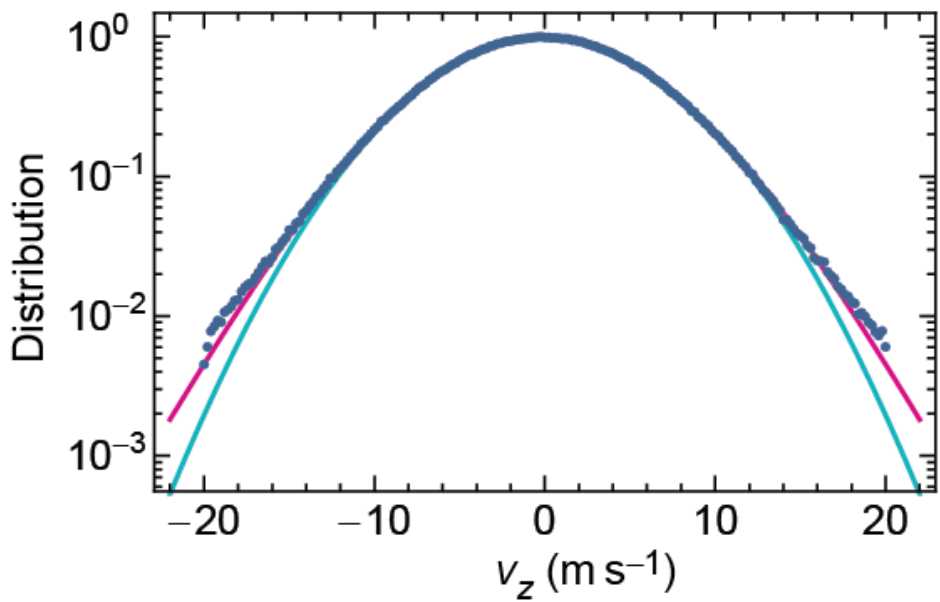

**Figure 3| Non-thermal velocity distributions.** When kinematic effects of (laser-induced) chemistry are included in MD simulations, the HD$^+$ z-velocity distribution exhibits non-thermal distributions with pronounced tails (blue data points). Indeed, a least-squares fit with a *q*-Gaussian (magenta curve, with fitted temperature *T*=11.00(3) mK and *q* = 1.070(3)) produces a visibly better result than a fit with a thermal (normal) distribution (turquoise curve, fit with *T*= 11.60(3) mK).

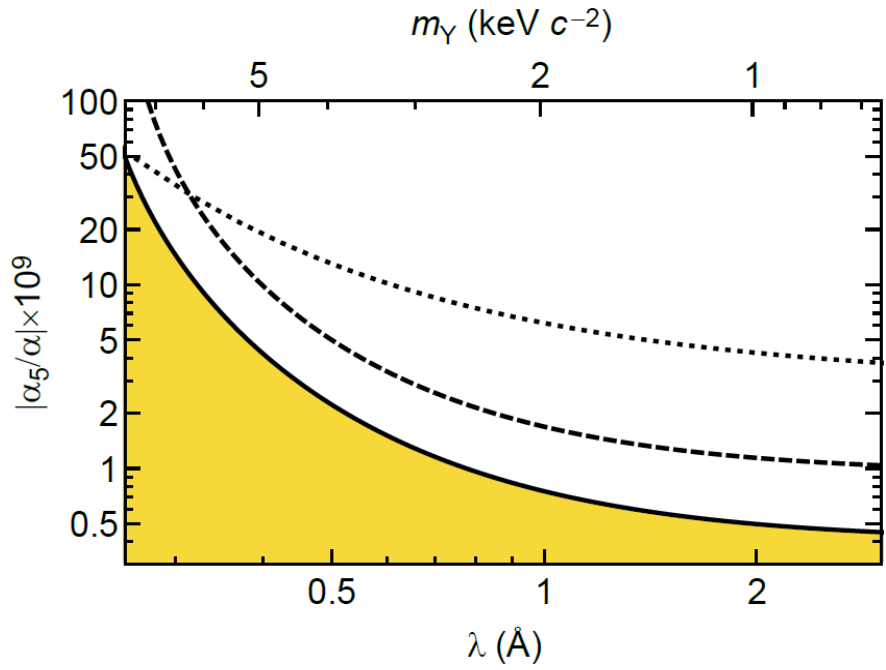

**Figure 4| Constraint on fifth forces between hadrons at the Ångstrom scale.** The high accuracy of our result and the good agreement with theory allow constraining the strength of hypothetical hadron-hadron interactions[5] with a characteristic range of the order of 1 Å, mediated through virtual particles with a mass of order 1 keV$c^{-2}$. The present result (solid curve) improves several times on the previous best constraints obtained from HD$^+$ (dashed curve) and neutral molecular hydrogen (D$_2$, dotted curve)[5], ruling out interactions with strengths $|\alpha_5/\alpha| > 5\times10^{-10}$ (90% confidence level). The yellow region remains unexplored by this experiment.

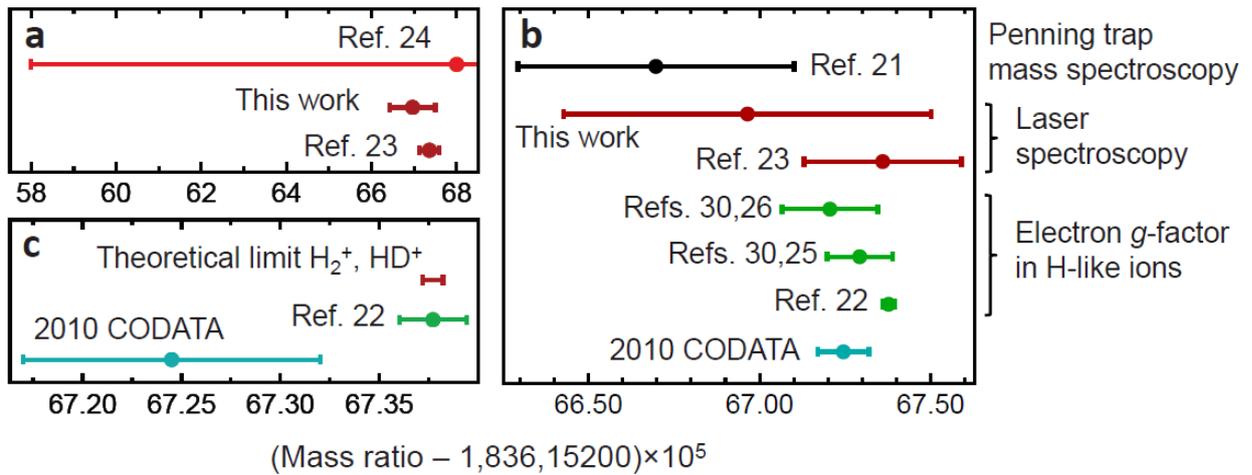

**Figure 5| Determination of the proton-electron mass ratio. a**, Comparison of values of $\mu$ obtained as a single parameter from laser spectroscopy. Shown are results from atomic hydrogen[24], atomic antiprotonic helium (which assumes charge, parity and time reversal invariance)[23], and molecular spectroscopy (this work). Error bars represent one standard deviation. **b,** Overview of the values included in the 2010 CODATA value of $\mu$, obtained through different methods, in comparison with this work. **c,** If experimental accuracy is improved beyond the ~$1\times10^{-11}$ relative uncertainty of state-of-the-art QED theory[1,2], $\mu$ could in principle be determined from the molecular hydrogen ion with a precision (red bar) which exceeds that of the current best $\mu$ determinations.

**Table 1| Systematic shifts and uncertainty budget.**

| Origin | Shift | Uncertainty | |
|---|---|---|---|
| | (MHz) | (MHz) | (p.p.b) |
| Resolution (statistical fit error) | 0 | 0.33 | 0.85 |
| Doppler effect due to chemistry | -0.25* | 0.25 | 0.66 |
| Ignoring population $L$=6 in rate equations | 0 | 0.032 | 0.083 |
| Doppler effect due to micromotion | -0.055* | 0.020 | 0.052 |
| Frequency measurement | 0 | 0.010 | 0.026 |
| BBR temperature | 0 | 0.005 | 0.013 |
| Zeeman effect | -0.0169 | 0.003 | 0.008 |
| Stark effect | -0.0013 | 0.0001 | 0.0004 |
| Electric-quadrupole shift | 0† | 0.0001 | 0.0003 |
| 2$^{nd}$-order Doppler effect | 0† | 0.000005 | 0.00001 |
| Total | -0.0182 | 0.41 | 1.1 |

Values labelled with an asterisk correspond to frequency shifts which are accounted for by the spectral fit function, rather than being subtracted from $\nu_{0,\text{fit}}$. Values labelled with a dagger are non-zero but negligibly small and therefore ignored here.